\documentclass{article}
\usepackage{amsmath, amssymb, amsfonts}
\title{Quantum potential and wave packet spreading}  
\author{Hristu Culetu, \\Ovidius University, Dept.of Physics and Electronics, \\ Mamaia Avenue 124, 900527 Constanta, Romania, \\e-mail : hculetu@yahoo.com}

\begin{document}
\numberwithin{equation}{section}
\pagenumbering{arabic}
\maketitle
\newcommand{\fv}{\boldsymbol{f}}
\newcommand{\tv}{\boldsymbol{t}}
\newcommand{\gv}{\boldsymbol{g}}
\newcommand{\OV}{\boldsymbol{O}}
\newcommand{\wv}{\boldsymbol{w}}
\newcommand{\WV}{\boldsymbol{W}}
\newcommand{\NV}{\boldsymbol{N}}
\newcommand{\hv}{\boldsymbol{h}}
\newcommand{\yv}{\boldsymbol{y}}
\newcommand{\RE}{\textrm{Re}}
\newcommand{\IM}{\textrm{Im}}
\newcommand{\rot}{\textrm{rot}}
\newcommand{\dv}{\boldsymbol{d}}
\newcommand{\grad}{\textrm{grad}}
\newcommand{\Tr}{\textrm{Tr}}
\newcommand{\ua}{\uparrow}
\newcommand{\da}{\downarrow}
\newcommand{\ct}{\textrm{const}}
\newcommand{\xv}{\boldsymbol{x}}
\newcommand{\mv}{\boldsymbol{m}}
\newcommand{\rv}{\boldsymbol{r}}
\newcommand{\kv}{\boldsymbol{k}}
\newcommand{\VE}{\boldsymbol{V}}
\newcommand{\sv}{\boldsymbol{s}}
\newcommand{\RV}{\boldsymbol{R}}
\newcommand{\pv}{\boldsymbol{p}}
\newcommand{\PV}{\boldsymbol{P}}
\newcommand{\EV}{\boldsymbol{E}}
\newcommand{\DV}{\boldsymbol{D}}
\newcommand{\BV}{\boldsymbol{B}}
\newcommand{\HV}{\boldsymbol{H}}
\newcommand{\MV}{\boldsymbol{M}}
\newcommand{\be}{\begin{equation}}
\newcommand{\ee}{\end{equation}}
\newcommand{\ba}{\begin{eqnarray}}
\newcommand{\ea}{\end{eqnarray}}
\newcommand{\bq}{\begin{eqnarray*}}
\newcommand{\eq}{\end{eqnarray*}}
\newcommand{\pa}{\partial}
\newcommand{\f}{\frac}
\newcommand{\FV}{\boldsymbol{F}}
\newcommand{\ve}{\boldsymbol{v}}
\newcommand{\AV}{\boldsymbol{A}}
\newcommand{\jv}{\boldsymbol{j}}
\newcommand{\LV}{\boldsymbol{L}}
\newcommand{\SV}{\boldsymbol{S}}
\newcommand{\av}{\boldsymbol{a}}
\newcommand{\qv}{\boldsymbol{q}}
\newcommand{\QV}{\boldsymbol{Q}}
\newcommand{\ev}{\boldsymbol{e}}
\newcommand{\uv}{\boldsymbol{u}}
\newcommand{\KV}{\boldsymbol{K}}
\newcommand{\ro}{\boldsymbol{\rho}}
\newcommand{\si}{\boldsymbol{\sigma}}
\newcommand{\thv}{\boldsymbol{\theta}}
\newcommand{\bv}{\boldsymbol{b}}
\newcommand{\JV}{\boldsymbol{J}}
\newcommand{\nv}{\boldsymbol{n}}
\newcommand{\lv}{\boldsymbol{l}}
\newcommand{\om}{\boldsymbol{\omega}}
\newcommand{\Om}{\boldsymbol{\Omega}}
\newcommand{\Piv}{\boldsymbol{\Pi}}
\newcommand{\UV}{\boldsymbol{U}}
\newcommand{\iv}{\boldsymbol{i}}
\newcommand{\nuv}{\boldsymbol{\nu}}
\newcommand{\muv}{\boldsymbol{\mu}}
\newcommand{\lm}{\boldsymbol{\lambda}}
\newcommand{\Lm}{\boldsymbol{\Lambda}}
\newcommand{\opsi}{\overline{\psi}}
\renewcommand{\tan}{\textrm{tg}}
\renewcommand{\cot}{\textrm{ctg}}
\renewcommand{\sinh}{\textrm{sh}}
\renewcommand{\cosh}{\textrm{ch}}
\renewcommand{\tanh}{\textrm{th}}
\renewcommand{\coth}{\textrm{cth}}

\begin{abstract}
The effects of the de Broglie-Bohm quantum potential on a test particle of mass $m$ are investigated in a conformally-flat geometry. A real, nonlinear, scalar field $\Psi$ is introduced and related directly to the conformal factor and to the effective mass $M(r,t)$ of the particle. The radial acceleration of a static observer in the conformally-flat metric is negative (the field is repulsive), and the corresponding proper acceleration resembles that one of the ''peak radius'' where the radial probability density has a global maximum during the wave packet spreading phenomenon. The timelike radial geodesics are computed in the conformally-flat spacetime in double-null coordinates and proves to be hyperbolae in the region $r>t$.
 \end{abstract}

 \section{Introduction}
The problem of reconciling Quantum Mechanics (QM) equations with the equations of classical physics remains unsolved. An alternative possibility may be Bohmian mechanics or the de Broglie-Bohm interpretation \cite{BHK}. A particle follows a definite path, determined not only by the local potential but also by the nonlocal feedback of the particle wave function on its own motion \cite{GRL}. Experiments have also shown that the de Broglie-Bohm trajectories can have a clear physical interpretation if full nonlocality is taken into account \cite{DM}

Gregori et al. \cite{GRL} develop a formalism which offers a reinterpretation of the classical Bohmian path by including the effect of Bohm's nonlocal potential on the particle motion, taken as an external field and introduced via conformal transformation on the line element. They also extended their model by means of a relativistic version of the Bohmian potential (see also \cite{SS, SD})
   \begin{equation}
	V_{B} = \frac{\hbar^{2}}{2m}\frac{\Box |\Psi|}{|\Psi|},
 \label{1.1}
 \end{equation}
where ''$\Box$'' stands for the d'Alembertian operator and $m$ is the mass of the particle. To get a positive definite ''effective'' mass $M$, Shojai and Shojai \cite{SS} suggested an expression of the form
   \begin{equation}
	M = me^{\frac{1}{2m^{2}}\frac{\Box |\Psi|}{|\Psi|}}.
 \label{1.2}
 \end{equation}
de Broglie \cite{LB} pointed out that the inclusion of quantum effects are completely equivalent to a conformal transformation of the spacetime metric. The authors of \cite{SS} developed the de Broglie result to ensure it satisfies the non-relativistic limit. Once one has the expression of $\Psi$, quantum effects on a classical particle can be calculated via the above recipe \cite{GRL}. 

  We follow in this paper a similar prescription using a real scalar field $\Psi$. In addition, our field satisfies a conformal-invariant equation, the self-interacting field potential being proportional with $\Psi^{4}$ \cite{DGB}. We further use that scalar field as a conformal factor of a conformally flat metric in spherical coordinates and found that the expression of the radial acceleration of a static observer resembles the acceleration of the ''peak'' radius where the radial probability density has a maximum, during wave packet spreading phenomenon \cite{GG}. The timelike radial geodesics in the conformally flat metric are computed in the both regions $r>t$ and $r<t$. They prove to be straightlines or hyperbolae.
	
	We use the positive signature for the line element and the geometrical units $\hbar = c =1$, unless otherwise specified. 
	
	\section{Conformal scalar field equation of motion}
Our main goal is to find the role played by de Broglie-Bohm quantum potential upon the behavior of a particle motion. Let us assume that our scalar field has the following expression
   \begin{equation}
	\Psi (r,t) = \frac{k}{\sqrt{r^{2}-t^{2}}}
 \label{2.1}
 \end{equation}
in spherical coordinates $r, t$ with $r>t$ and $k$-a positive dimensionless constant. If one calculates the d'Alembertian of this field in Minkowski spacetime, one finds that
   \begin{equation}
	\Box \Psi = -\frac{\partial^{2}\Psi}{\partial t^{2}} + \frac{\partial^{2}\Psi}{\partial r^{2}} + \frac{2}{r} \frac{\partial \Psi}{\partial r} = -k\Psi^{3}
 \label{2.2}
 \end{equation}
In other words, $\Psi$ satisfies the equation of motion
   \begin{equation}
	\Box \Psi + k\Psi^{3} = 0, 
 \label{2.3}
 \end{equation}
where $\Psi$ is an inverse distance and $k$ - the coupling constant. It is well-known that the nonlinear equation (2.3) is conformally-invariant \cite{DGB} in 4-dimensions. In addition, $\Psi$ is massless (no mass term appears in (2.3)). Moreover, it explodes on the light-cone $r = t$.
As de Broglie has stated \cite{LB}, the inclusion of quantum effects is equivalent with a conformal transformation on the metric, with a conformal factor depending on the ''effective'' mass $M(r,t)$. Therefore, our starting geometry is a conformally-flat one
   \begin{equation}
	  ds^{2} = e^{\frac{b^{2}}{r^{2}-t^{2}}}(-dt^{2} + dr^{2} + r^{2} d \Omega^{2}), ~~~~r>t,~~~t>0,
 \label{2.4}
 \end{equation}
where $b$ is taken as the Compton wavelength of the particle of mass $m$ and $d \Omega^{2}$ stands for the metric on the unit 2-sphere. One notices also that the conformal factor directly depends on $\Psi$ and the mass $m$. The above line-element is curved, the scalar curvature and the Kretschmann scalar are finite everywhere and vanish on the surface $r = t$ and at $r = 0$. Another quantity that deserves being computed is the acceleration of a static observer in the spacetime (2.4). The velocity 4-vector will be
   \begin{equation}
	  u^{a} = \left(e^{\frac{-b^{2}}{2(r^{2}-t^{2})}}, 0, 0, 0\right),
 \label{2.5}
 \end{equation}
where $a$ runs from 0 to 3 and $u^{a}u_{a} = -1$. The 4-acceleration $a^{b} = u^{a}\nabla_{a}u^{b}$ looks like
   \begin{equation}
	  a^{b} = \left(0, -\frac{b^{2}r}{(r^{2}-t^{2})^{2}} e^{-\frac{b^{2}}{r^{2}-t^{2}}}, 0, 0\right).
 \label{2.6}
 \end{equation}
It is worth noting that the radial component $a^{r}$ is always negative and vanishes at $r = t$ and $r = 0$ (which imposes $t = 0$, too, because $t<r$). The fact that $a^{r}<0$ suggests a repulsive field (a static observer has to push himself to the origin $r = 0$ for maintaining the same radial position). The proper acceleration obtained from (2.6) is 
   \begin{equation}
	 A \equiv \sqrt{a^{b}a_{b}} = \frac{b^{2}r}{(r^{2}-t^{2})^{2}} e^{-\frac{b^{2}}{2(r^{2}-t^{2})}}.
 \label{2.7}
 \end{equation}
Two remarks are in order here: firstly, $A$ is zero when $\hbar = 0$ and, secondly, due to the Compton wavelength $b$, the larger the mass of the particle finding in the field $\Psi$, the smaller the acceleration $A$ is. As we see, the macroscopic masses are not influenced too much by the field. 

  For $r>>t$ and $r>>b$, (2.7) gives us in first order approximation in $t/r$ and $b/r$
	   \begin{equation}
	 A = \frac{b^{2}}{r^{3}} e^{-\frac{b^{2}}{2r^{2}}} \approx \frac{b^{2}}{r^{3}} = \frac{\hbar^{2}}{m^{2}r^{3}}
 \label{2.8}
 \end{equation}
where we introduced the fundamental constants in the last ratio. For example, at $r = 10^{2}b$ and $m = m_{e} = 10^{-27}g$, we get $A \approx 10^{15} cm/s^{2}$. But that is valid for $t<<r$, which means $t<<10^{-18}s$. In other words, the duration of the measurement has to be much less than $10^{-18}$s for to obtain the above value of $A$. If $r = 1 cm$, the acceleration becomes $A \approx 1 cm/s^{2}$, if measured in a time $t<<10^{-10}s$. Moreover, $A$ from (2.8) (including the exponential factor), as a function of $r$, has a maximum value at $r = b\sqrt{3}$, with $A_{max} = (3\sqrt{3}/e\sqrt{e})(mc^{3}/\hbar$) and is vanishing when $r \rightarrow 0$ or $r \rightarrow \infty$. It is worth observing that the limit $r \rightarrow 0$ is appropriate only when $m$ is considered a point particle. In QM it is extended over a Compton wavelength $b$, not having a precise location due to the Heisenberg Principle. Even though the $\Psi$-field is not quantized, it depends on Planck's constant so that our approach is semiclassical. 
  We point out that the shear and vorticity tensors are vanishing in the spacetime (2.4) when the velocity vector field is given by (2.5). In addition, the Shojai and Shojai \cite{SS} effective mass becomes in our situation
 \begin{equation}
	 M(r,t) = m e^{-\frac{b^{2}}{2(r^{2}-t^{2})}} = m e^{-\frac{1}{2m^{2}(r^{2}-t^{2})}} 
 \label{2.9}
 \end{equation}
(we chose above $k = 1$). One sees from (2.9) that $M$ is zero on the surface $r = t$, $M = m$ when the Minkowski interval $\sqrt{r^{2}-t^{2}}$ is much larger than Compton's wavelength $b = 1/m$ and $M$ vanishes when $m \rightarrow 0$. All these properties of $M$ seems reasonable. 
 From now on we consider $m$ a test particle ''guided'' by the nonlinear field $\Psi$. However, $m$ appears in the metric (2.4) through the length $b$, which has been introduced for dimensional reasons. Its behavior is generated by the nonlinearity of Eq. (2.3). One observes the conformal factor of (2.4) has no effect (i.e., $M \approx m$) when the particle is macroscopic, say a billiard ball, and when our observer is not located very close to the light cone $r = t$. 

\section{Wave packet spreading analogy}
 Starting with the exact solution of the time dependent Schrodinger equation for a free particle, the authors of \cite{GG} wrote down the global maximum of the radial probability density
  \begin{equation}
	\ro (r,t) = \frac{4r^{2}}{\sqrt{\pi}\sigma^{3}} \frac{e^{-\frac{r^{2}}{\sigma^{2}\left(1 + \frac{t^{2}}{m^{2}\sigma^{4}}\right)}}}{\left(1 + \frac{t^{2}}{m^{2}\sigma^{4}}\right)^{3/2}}
 \label{3.1}
 \end{equation}
 (see also \cite{BLM}), where $\sigma$ is the initial width of the Gaussian wave packet. Eq. (3.1) yields
  \begin{equation}
	\frac{\partial \ro}{\partial r} = f(t,\sigma) \frac{\partial}{\partial r}\left[r^{2}e^{-\frac{r^{2}}{g(t,\sigma)}}\right] = 0,
 \label{3.2}
 \end{equation}
where the significance of the functions $f$ and $g$ is clear from (3.1). One obtains
  \begin{equation}
	1 - \frac{r^{2}}{g(t,\sigma)} = 0,
 \label{3.3}
 \end{equation}
whence \cite{GG}
  \begin{equation}
	r_{p} = \sigma \sqrt{1 + \frac{t^{2}}{m^{2}\sigma^{4}}}.
 \label{3.4}
 \end{equation}
The index ''p'' stands for ''peak'' and $r_{p}$ represents the location of the maximum of the probability density \cite{SC}. It ''accelerates'' outward at the rate $d^{2}r_{p}/dt^{2} = 1/(m^{2}r_{p}^{3})$. Eq. (3.1) expresses the hyperbolic expansion (wave packet spreading) of the free particle. 

From (3.4) one finds that the spreading velocity is given by
  \begin{equation}
	\frac{dr_{p}}{dt} = \frac{1}{m \sigma} \frac{t}{\sqrt{t^{2} + \left(\frac{\sigma}{v_{\infty}}\right)^{2}}}
 \end{equation}
where $v_{\infty} = 1/m\sigma$ is the velocity at infinity or when $t>>m\sigma^{2}$. Consequently, the maximum velocity is related to the initial width of the wave packet, which becomes the velocity of light if $\sigma$ is considered as the Compton wavelength associated to the mass $m$.

We compare now the previous ''acceleration'' with $A$ from (2.8). They have the same mathematical form in the approximation used to get (2.8). On the grounds of the above analogy, we assume that $A$ has the meaning of an outward acceleration generated by the wave packet expansion (we saw that the radial acceleration $a^{r}$ is repulsive). Put it differently, due to the conformal factor from (2.4), the field $\Psi$ spreads the mass $m$ that becomes the effective mass $M(r,t)$, copying the wave packet dispersion phenomenon. We notice also that Newton's constant $G$ appears nowhere in our equations: the only influence of gravity is through the curved metric (2.4), whose conformal factor vanishes when $m \rightarrow 0$. A feedback mechanism arises here: the particle finds itself in a field $\Psi$ which, in turn, generates the conformal factor which contains its mass $m$.

\section{Timelike radial geodesics ($r>t$)} 
Let us investigate now the geodesics of a freely falling observer in the geometry (2.4). For that purpose it is easier to use null coordinates $v = r+t$ and $u = r-t$, with $v>0,~u>0$. Eq. (2.4) in these new coordinates acquires the form
   \begin{equation}
	  ds^{2} = e^{\frac{b^{2}}{uv}} \left[du~dv + \left(\frac{u+v}{2}\right)^{2} d \Omega^{2}\right], 
 \label{4.1}
 \end{equation}
with
   \begin{equation}
	  \dot{u} \dot{v} = - e^{-\frac{b^{2}}{uv}}  
 \label{4.2}
 \end{equation}
for the radial geodesics ($\dot{u} = du/d\tau,~\dot{v} = dv/d\tau$, where $\tau$ is the proper time). To find the geodesic curves, one begins with the Lagrangean
   \begin{equation}
	 L = - e^{\frac{b^{2}}{uv}}  \dot{u} \dot{v}.  
 \label{4.3}
 \end{equation}
From the Euler-Lagrange equations
 \begin{equation}
 \frac{\partial L}{\partial v} - \frac{d}{d\tau}\left( \frac{\partial L}{\partial \dot{v}}\right) = 0, ~~~ \frac{\partial L}{\partial u} - \frac{d}{d\tau}\left( \frac{\partial L}{\partial \dot{u}}\right) = 0,
 \label{4.4}
 \end{equation}
one finds, using (4.2), that
 \begin{equation}
\frac{b^{2}}{uv^{2}} - \frac{d}{d\tau}\left(\dot{u} e^{\frac{b^{2}}{uv}} \right) = 0,~~~\frac{b^{2}}{u^{2}v} - \frac{d}{d\tau}\left(\dot{v} e^{\frac{b^{2}}{uv}} \right) = 0.
 \label{4.5}
 \end{equation}
Operating on the last equations, we get
  \begin{equation}
 \frac{d}{d\tau}\left(\frac{u}{\dot{u}} \right) - \frac{d}{d\tau}\left(\frac{v}{\dot{v}} \right) = 0,
 \label{4.6}
 \end{equation}
which gives us
  \begin{equation}
\frac{u}{\dot{u}} = \frac{v}{\dot{v}} + \alpha ,
 \label{4.7}
 \end{equation}
where $\alpha$ is a constant of integration. We have two cases here:\\
i) $\alpha$ = 0.\\
The above choice leads to $v = \kappa_{1}u~(\kappa_{1} > 1)$ and, therefore $r = (\kappa_{1} + 1)/(\kappa_{1} - 1) t > t$, which is a tachyonic equation. The fact that this is impossible could be seen from (4.2), because $\dot{u} \dot{v} = \dot{r}^{2} - \dot{t}^{2} < 0$, in contradiction with the above equation for $r(t)$. \\
ii) $\alpha \neq 0$\\
We try a solution of (4.7) of the form
  \begin{equation}
	u~v = \beta,
 \label{4.8}
 \end{equation}
where $\beta$ is a positive constant. From (4.7) and (4.8) one easily finds that
  \begin{equation}
	u(\tau) = \gamma e^{\frac{2\tau}{\alpha}},~~~v(\tau) = \frac{\beta}{\gamma} e^{-\frac{2\tau}{\alpha}}.
 \label{4.9}
 \end{equation}
We choose $\gamma = \sqrt{\beta}$ for to get $t = 0$ at $\tau = 0$. From (4.9) one obtains
  \begin{equation}
	r(\tau) = \frac{\sqrt{\beta}}{2} \left(e^{\frac{2\tau}{\alpha}} + e^{-\frac{2\tau}{\alpha}}\right),~~~t(\tau) = \frac{\sqrt{\beta}}{2} \left(e^{-\frac{2\tau}{\alpha}} - e^{\frac{2\tau}{\alpha}}\right).
 \label{4.10}
 \end{equation}
It is clear that we have to impose $\alpha < 0$ as long as $t > 0$. In terms of the hyperbolic functions, (4.10) appears as 
  \begin{equation}
	r(\tau) = \sqrt{\beta} cosh~\frac{2\tau}{\alpha} ,~~~t(\tau) = -\sqrt{\beta} sinh~\frac{2\tau}{\alpha},
 \label{4.11}
 \end{equation}
which gives us 
  \begin{equation}
	r^{2} - t^{2} = \beta .
 \label{4.12}
 \end{equation}
From (4.2) one obtains
  \begin{equation}
	\alpha^{2} = 4\beta e^{\frac{b^{2}}{\beta}}.
 \label{4.13}
 \end{equation}
In addition, Eqs. (4.5) have to be obeyed. That gives us $\beta = b^{2}$ and, from (4.13), we have $\alpha = -2b\sqrt{e}$. Consequently, $\beta$ is not arbitrary, taking a unique value. Therefore, radial geodesics are given by
  \begin{equation}
	r(t) = \sqrt{t^{2} + b^{2}}.
 \label{4.14}
 \end{equation}
In other words, the metric (2.4) generates, through the quantum potential $\Psi$, a hyperbolic expansion of the wave packet. $r(t)$ from (4.14) is interpreted as the location at time $t$ of the boundary of the wave packet, viewed as a wave front. Compared to (3.4) one observes that the initial width of the wave packet is $\sigma = b$, so that $v_{\infty}$ from (3.5) equals the velocity of light due to the massless character of the nonlinear field $\Psi$.

\section{Timelike radial geodesics ($t>r$)}
In the region $t>r$ we take the scalar field as
  \begin{equation}
		\bar{\Psi} (r,t) = \frac{k}{\sqrt{t^{2}-r^{2}}}
 \label{5.1}
 \end{equation}
instead of  $\Psi$ from (2.1). On the other hand, one obtains now
   \begin{equation}
	\Box \bar{\Psi} - k \bar{\Psi}^{3} = 0, 
 \label{5.2}
 \end{equation}
and, taking again $k = 1$ the nonlinear term and the first one containing the time derivative will have the same sign. 

  The behavior of the proper acceleration 
	   \begin{equation}
	 \bar{A} = \frac{b^{2}r}{(t^{2}-r^{2})^{2}} e^{-\frac{b^{2}}{2(t^{2}-r^{2})}}
 \label{5.3}
 \end{equation}
is now different compared to $A$ from (2.8) (however, again $a^{r}<0$). At the initial time $t = 0$ we have to set simultaneously $r = 0$, since $t\geq r$. In that case, $\bar{A}$ from (5.3) vanishes, as can be seen directly by performing the limit $t^{2}-r^{2} \rightarrow 0$. With $t>>r,~t>>b$ in (5.3), we have, in a first order approximation
 	\begin{equation}
	 \bar{A} \approx \frac{b^{2}r}{t^{4}} = \frac{\hbar^{2}r}{m^{2}c^{4}t^{4}},
 \label{5.4}
 \end{equation}
if one keeps track of the fundamental constants. Written differently, (5.4) becomes
 	\begin{equation}
	 \bar{A} = \omega^{2}~ r,
 \label{5.5}
 \end{equation}
where $\omega = 1/mt^{2}$ is an angular frequency. Formula (5.5) resembles the centrifugal acceleration of a rotating body with a decreasing $\omega$. As an example, let us consider an electron with $m_{e} \approx 10^{-27}g$ and take the special value $r = b$. In that case, we must have $t>>10^{-21}s$. Take, say, $t = 10^{-14} s$. One obtains $\bar{A} \approx 10^{7} cm/s^{2}$ and $\omega \approx 10^{8} s^{-1}$. Let us observing that, when the mass $m$ is macroscopic, say $m = 1g$, one obtains $b \approx 10^{-37} cm$. At $r = 10 cm$ from its center of mass and with, say, $ct = 10^{4} cm >>r$, we get $\bar{A} \approx 10^{-69} cm/s^{2}$ and $\omega \approx 10^{-35} s^{-1}$, completely negligible values. 

We shall find now the geodesics in the spacetime
   \begin{equation}
	  ds^{2} = e^{\frac{b^{2}}{t^{2}-r^{2}}}(-dt^{2} + dr^{2} + r^{2} d \Omega^{2}), ~~~~t>r
 \label{5.6}
 \end{equation}
using the null coordinates $\bar{v} = t-r,~\bar{u} = t+r$, with $\bar{u},~\bar{v}>0,~\bar{v}<\bar{u}$. The metric (5.6) acquires the form
   \begin{equation}
	  ds^{2} = e^{\frac{b^{2}}{\bar{u} \bar{v}}} \left[-d\bar{u}~d\bar{v} + \left(\frac{\bar{u} - \bar{v}}{2}\right)^{2} d \Omega^{2}\right], 
 \label{5.7}
 \end{equation}
with
   \begin{equation}
	  \dot{\bar{u}} \dot{\bar{v}} = e^{-\frac{b^{2}}{\bar{u} \bar{v}}}. 
 \label{5.8}
 \end{equation}
Following the same steps as in Sec.4, one finds that
  \begin{equation}
\frac{\bar{u}}{\dot{\bar{u}}} = \frac{\bar{v}}{\dot{\bar{v}}} + \bar{\alpha} .
 \label{5.9}
 \end{equation}
i)~$\bar{\alpha} = 0$\\
We obtain $\bar{v} = \kappa_{2} \bar{u},~0<\kappa_{2}<1$, whence
  \begin{equation}
	r = \frac{1 - \kappa_{2}} {1 + \kappa_{2}}~ t,
 \label{5.10}
 \end{equation}
which is a straight line. One observes the solution (5.10) is independent on the value of $b$.

Our next purpose is finding $\bar{u},~\bar{v}$ in terms of $\tau$. We have $\dot{\bar{v}} = \kappa_{2} \dot{\bar{u}}$ and from (5.8)
   \begin{equation}
	  \sqrt{\kappa_{2}} \dot{\bar{u}}  = e^{-\frac{b^{2}}{2\kappa_{2}\bar{u}^{2}}}  
 \label{5.11}
 \end{equation}
With the new variable $y^{2} = b^{2}/(2\kappa_{2} \bar{u}^{2})$, Eq. (5.11) yields
   \begin{equation}
	\frac{e^{y^{2}}}{y^{2}} dy = -\frac{\sqrt{2}}{b} d\tau ,
 \label{5.12}
 \end{equation}
which cannot be exactly integrated. A series expansion leads to
   \begin{equation}
	\int{\frac{e^{y^{2}}}{y^{2}} dy} = -\frac{1}{y} + y + \frac{y^{3}}{3\cdot 2!} + \frac{y^{4}}{4\cdot 3!} +...,
 \label{5.13}
 \end{equation}
whence
   \begin{equation}
	\frac{\sqrt{2\kappa_{2}}}{b} \bar{u} - \frac{b}{\sqrt{2\kappa_{2}} \bar{u}} - \frac{b^{3}}{\sqrt{2\kappa_{2}}~ \bar{u}^{3} \cdot 3 \cdot 2!} - ... = \frac{\sqrt{2}}{b} \tau .
 \label{5.14}
 \end{equation}
A similar equation for $\bar{v}(\tau)$ is obtained, plugging $\bar{v} = \kappa_{2} \bar{u}$ in (5.14).\\
ii)~$\bar{\alpha} \neq 0$\\
Looking for a solution of the form $\bar{u} \bar{v} = \bar{\beta}>0$ and using (5.8) one obtains
  \begin{equation}
	\bar{u}(\tau) = \sqrt{\bar{\beta}}~e^{\frac{2\tau}{\bar{\alpha}}},~~~\bar{v}(\tau) = \sqrt{\bar{\beta}} ~e^{-\frac{2\tau}{\bar{\alpha}}} 
 \label{5.15}
 \end{equation}
($\bar{\alpha}>0$), which gives us
  \begin{equation}
	t^{2} - r^{2} = \bar{\beta}>0,
 \label{5.16}
 \end{equation}
which is a tachyonic equation. As in the previous case ($r>t$), one arises a similar inconsistence: $\dot{\bar{u}} \dot{\bar{v}} = -\frac{4}{\bar{\alpha}^{2}}uv<0$, in contradiction with (5.8). 

To summarize, we found that, for $t>r$, the only geodesic is a straightline but for $r>t$ the radial geodesic proves to be a hyperbola.

\section{Conclusions}
The evolution of a quantum particle in a real, nonlinear scalar field $\Psi (r,t)$ was investigated in a conformally flat spacetime. The de Broglie-Bohm quantum potential is related to the effective mass of the test particle whose acceleration is interpretated as that of the radial distance where the probability density has an extremum, during the wave packet spreading phenomenon for a free particle. The field $\Psi$ satisfies a conformally invariant equation, with a term proportional to $\Psi^{3}$. The effect of the quantum potential is negligible when the test particle is macroscopic. The radial timelike geodesics were computed both for $r>t$ and for $t>r$. In the region $r>t$ there is only one radial geodesic $r(t) = \sqrt{t^{2} + b^{2}}$ and those from the region $t>r$ are straightlines.


\begin{thebibliography} {10}

\bibitem{BHK}
D. Bohm, B. Hiley and P. Kaloyerou, Phys. Rep. 144, 321 (1987).
\bibitem{GRL}
G. Gregori, B. Reville and B. Larder, arXiv: 1904.12388.
\bibitem {DM}
D. H. Mahler et al., Sci. Adv. e1501466 (2016).
\bibitem{SS}
A. Shojai and S. Shojai, Phys. Scripta 64, 413 (2006). 
\bibitem{SD}
S. Das, Phys. Rev. D89, 084068 (2014); arXiv: 1311.6539.
\bibitem{LB}
L. de Broglie, Nonlinear wave mechanics. A causal interpretation (Elsevier, Amsterdam, 1960).
\bibitem {DGB}
M. Dabrovski, J. Garecki and D. Blaschke, Annalen Phys. (Berlin), 18, 13 (2009); arXiv: 0806.2683.
\bibitem {GG}
D. Giulini and A. Gro$\beta$ardt, Class. Quantum Grav. 28, 195026 (2011); arXiv: 1105.1921.
\bibitem{SC}
S. Carlip, Class. Quantum Grav. 25, 154010 (2008); arXiv: 0803.3456.
\bibitem{BLM}
L. Buoninfante, G. Lambiase and A. Mazumdar, Eur. Phys. J. C78, no.1, 73 (2018); arXiv: 1709.09263.



\end{thebibliography}
\end{document}